\documentclass[a4paper,11pt]{article}
\usepackage{jcappub}
\usepackage{lineno}
\usepackage[utf8,latin1]{inputenc}
\usepackage{graphicx}
\usepackage{dcolumn}
\usepackage[dvipsnames]{xcolor}
\usepackage[T1]{fontenc}

\usepackage{mathrsfs}  
\usepackage{cases}
\usepackage{bm}
\usepackage{academicons}
\usepackage{mathtools, nccmath}
\usepackage{fancyhdr}
\usepackage{tikz,xcolor}

\usepackage{tensor}
\usepackage[normalem]{ulem}
\usepackage{lipsum}
\usepackage{soul}
\usepackage{cancel}
\usepackage{stackengine,scalerel}
\usepackage{mathabx}
\usepackage{hyperref}
\usepackage{tabularx}
\usepackage[justification=raggedright]{caption}
\usepackage{placeins}
\hypersetup{colorlinks, linkcolor={red},citecolor={blue},urlcolor={blue}}

\definecolor{lime}{HTML}{A6CE39}
\DeclareRobustCommand{\orcidicon}{
	\begin{tikzpicture}
	\draw[lime, fill=lime] (0,0) 
	circle [radius=0.16] 
	node[white] {{\fontfamily{qag}\selectfont \tiny ID}};
	\draw[white, fill=white] (-0.0625,0.095) 
	circle [radius=0.007];
	\end{tikzpicture}
	\hspace{-2mm}
}

\newcommand{\bea}{\begin{eqnarray}}
\newcommand{\ena}{\end{eqnarray}}
\newcommand{\beann}{\begin{eqnarray*}}
\newcommand{\enann}{\end{eqnarray*}}

\fancypagestyle{plain}{%
  \fancyhf{}
  \fancyfoot[C]{\iffloatpage{}{\thepage}}
  }
\foreach \x in {A, ..., Z}{%
	\expandafter\xdef\csname orcid\x\endcsname{\noexpand\href{https://orcid.org/\csname orcidauthor\x\endcsname}{\noexpand\orcidicon}}
}

\begin{document}

\title{Signatures of modified gravity from the gravitational Aharonov-Bohm effect }

\author[a]{Kimet Jusufi\orcidA{}}
\author[b]{ Abdelrahman Yasser\orcidB{}}
\author[c]{Emmanuele Battista\orcidD{}}
\author[d,e,f]{Nader Inan \orcidC{}}

\affiliation[a]{\footnotesize{Physics Department, University of Tetova,
Ilinden Street nn, 1200, Tetovo, North Macedonia}}
\affiliation[b]{\footnotesize{Department of Physics, Faculty of Science, Cairo University, Giza 12613, Egypt}}
\affiliation[c]{\footnotesize{Istituto Nazionale di Fisica Nucleare, Laboratori Nazionali di Frascati, 00044 Frascati, Italy}}
\affiliation[d]{\footnotesize{Clovis Community College, 10309 N. Willow, Fresno, CA 93730 USA}}
\affiliation[e]{\footnotesize{Department of Physics, California State University Fresno, Fresno, CA 93740-8031, USA}}
\affiliation[f]{\footnotesize{University of California, Merced, School of Natural Sciences, P.O. Box 2039,
Merced, CA 95344, USA}}

\emailAdd{kimet.jusufi@unite.edu.mk} 
\emailAdd{ayasser@sci.cu.edu.eg}
\emailAdd{ebattista@lnf.infn.it,emmanuelebattista@gmail.com}
\emailAdd{ninan@ucmerced.edu }

\abstract{To date, no observational confirmation of dark matter particles has been found. In this paper, we put forward an alternative approach to inferring   evidence for dark matter through modified gravity, without invoking fundamental dark matter particles. Specifically, we explore the possibility of extracting signatures of Kaluza-Klein  gravity through the gravitational Aharonov-Bohm effect. Kaluza-Klein theory has recently been proposed as an alternative to the dark sector, and predicts a tower of particles, including spin-0 and spin-1 gravitons alongside the usual spin-2 graviton, which can gravitationally couple to matter. We thus analyze a quantum system in free fall around a gravitating body  in the presence of a modified Yukawa-like gravitational potential, and determine the gravitational phase induced by the additional degrees of freedom introduced by the Kaluza-Klein  model. Our results reveal that, in addition to the usual result from General Relativity, the quantum wave function of the system exhibits an additional effect: a splitting of  the energy levels with a new quantum number due to the extra vector gravitational degrees of freedom. The energy splitting difference between general relativity and Kaluza-Klein gravity is found to be of the order of meV for an atomic system and eV for a nuclear system. Similar values  also arise in generic modified gravity models and  can be feasibly  tested in the future. Numerical estimates for the graviton mass are also provided, and potential imprints on gravitational waves  are  mentioned. }

\maketitle
\flushbottom

\section{Introduction}

The Aharonov-Bohm (AB) effect is a  phenomenon that brings to light  the non-trivial topology of  quantum vacuum in gauge theories. First predicted  in 1959  \cite{AB}, it shows that when charged particles, such as electrons, traverse a region with a nonvanishing  electromagnetic potential but zero magnetic field, their quantum-mechanical wavefunction acquires a phase shift. This effect is typically  detected experimentally as a shift in the interference pattern in a double-slit setup \cite{Ryder1985} and has been  verified in a variety of systems \cite{Chambers1960,Batelaan2009,Gillot2013}.  

Recently, there has been interest in extending the AB effect to gravitational interactions, with particular attention to its scalar-gravitational manifestation. In these settings, the AB phenomenon arises from the proper time differences between freely falling, nonlocal trajectories, induced by the spacetime curvature. Such an effect was recently observed  by measuring the phase shift between two matter waves that traveled along different paths in a gravitational potential,   proving that gravitational potentials  produce AB phase shifts analogous to those of electromagnetic potentials \cite{Overstreet}. Moreover, the work of \cite{Chiao:2023ezj} has investigated the gravitational AB effect by considering a quantum system (such as an atomic or nuclear system) in a satellite orbiting the Earth,  exploring how a time-dependent gravitational potential influences the quantum states. The gravitational phase shift  leads to an oscillatory term in the energy spectrum, similar to previous work on the scalar-electric AB effect \cite{Chiao2022}. Interestingly, the scalar AB effect analyzed in Refs. \cite{Chiao:2023ezj,Chiao2022} induces side bands in the energy levels of the quantum system (observed via spectrometry), in contrast to an interference fringe shift seen in the usual vector AB effect (via interferometry).

General Relativity (GR) has been extensively tested and validated in  weak-field, low-velocity, and
linear gravity regimes \cite{Voisin2020,Will2014,Wex2014,Weisberg2004}. Furthermore, the recent detection of gravitational waves  from coalescing compact binaries provided the unprecedented opportunity to probe GR (as well as extended theories of gravity, see e.g. Refs. \cite{Trestini2024,DeFalco2024,Diedrichs2023,Bernard2022,Shiralilou2021,Battista2021}) also in the highly nonlinear, strong-field domain \cite{Blanchet2013,LIGOScientific2016a,LIGOScientific2018,LIGOScientific2019b,LIGOScientific2020,LIGOScientific2021b,Ghosh2022}. Beyond gravitational wave observations, the  recent striking images released by the Event Horizon Telescope collaboration \cite{EventHorizonTelescope:2020qrl,EventHorizonTelescope:2021srq,EventHorizonTelescope:2022wkp} have enabled a unique inspection of the nature of black holes as well as the behavior of matter in near-horizon regions \cite{Vagnozzi:2022moj,Bambi:2019tjh,Cunha:2019ikd}. Another crucial source for strong-gravity analyses comes from the long-term timing observations of binary pulsars, which allow for precision tests of GR predictions regarding strongly self-gravitating bodies, including effects of retardation and aberrational light bending \cite{Kramer2021}. However, although it accurately describes gravitational interactions in many contexts, GR   fails to account for some low-energy cosmological phenomena \cite{Copeland2006,Bertone2016,Salucci2020}. The examination of the galaxy rotation curves suggests the presence of additional unseen mass, thereby leading to the hypothesis of dark matter. Similarly, the accelerated expansion of the universe, inferred from the observation of distant supernovae, points to a mysterious force termed dark energy. Recent studies have shown that galaxies in the early Universe appear to have grown too large too quickly. On one hand, this poses a challenge for the $\Lambda$CDM model; on the other hand, it suggests a possible resolution to the structure formation paradigm through modified gravity \cite{McGaugh:2024yiz}. These issues indicate that  GR   may be incomplete at cosmological scales and   have led to the exploration of modified gravity theories (see e.g. Refs. \cite{Nojiri2010,Capozziello:2011et,Clifton2011} for a review), including those involving Yukawa-like potentials \cite{Jusufi:2024utf,Moffat:2005si,Aoki:2016zgp,Amendola:2019laa,Jusufi:2024ifp,J1,J2,J3,J4,J5}. Inspired by quantum field theory, Yukawa-like potentials introduce a gravitational force that decays exponentially with distance, thus diverging from the ordinary inverse-square law. Such modifications could explain galactic rotation curves without requiring dark matter and offer alternative perspectives on dark energy and cosmic expansion, potentially minimizing the need for exotic components in cosmology.

In this paper, we explore the possibility of applying the AB gravitational effect to Kaluza-Klein (KK) theory, which has  recently been proposed as an alternative to the dark sector \cite{Jusufi:2024utf}. This model introduces a generalized gravitational potential combining Yukawa and Newtonian terms, and brings in additional degrees of freedom, namely spin-0 and spin-1 gravitons, which represent  the gravitational analogs of  gauge fields. Furthermore, there is a complex scalar field $\tilde{\phi}$, which is minimally coupled to the gauge field $A_\mu$ with gauge coupling constant $g$. The structure of KK pattern is reminiscent of ultralight axion  models where a light scalar field acts as dark energy by modifying gravity over cosmological scales. Recent cosmological studies have shown that ultralight axions with masses less than  approximately $ 10^{-32} \, \text{eV}$ can contribute significantly to dark energy  \cite{Hlozek:2014lca}. This suggests that additional scalar degrees of freedom in modified gravity theories could provide an alternative explanation for cosmic acceleration. 

The main objective of the paper is to identify potential observational signatures arising from the gravitational AB effect within the context of KK gravity.  As we will demonstrate, this framework leads to intriguing results, such as shifts in the energy levels and the phase of a quantum system. These findings can potentially open up a new window for testing modified gravity theories using the gravitational AB effect. 

Recent studies have explored the gravitational AB effect as a tool to probe new physics. For instance, Ref. \cite{Nair:2024} examined how a charged graviton would experience an AB phase shift due to intergalactic magnetic fields, leading to new constraints from gravitational wave observations. In contrast, this work investigates the AB effect in a purely gravitational context, arising from modifications to the gravitational potential in extra dimensions.

The plan of the paper is as follows. After reviewing KK gravity theory in Sec. \ref{Sec:Yukawa potential}, we turn in Sec. \ref{Sec:Gravitational-AB-phase} to the key focus of this paper: studying the gravitational AB effect in a quantum system in free fall around a gravitating body (e.g., a satellite orbiting Earth). Then, in Sec. \ref{Sec:Energy-phase-shift}, we present numerical estimations on the shift in energy levels and the phase of the system, and provide some values of the graviton mass. Afterwards, in Sec. \ref{Sec:AB-effect-other-theories}, we investigate the AB effect in other modified gravity theories. Finally, we draw our conclusions  in Sec. \ref{Sec:Conclusions}. 

Throughout the paper, we use signature $(-+++)$ for the four-dimensional spacetime metric, and set $c=1$, unless otherwise specified. 

\section{Review of Kaluza-Klein gravity}\label{Sec:Yukawa potential}

In this section, we provide a review of KK gravity, closely following the framework outlined in Ref.~\cite{Jusufi:2024utf}. As we will see, we deal with an alternative interpretation of KK theory that does not involve electromagnetism. 

For simplicity, we consider a  KK model in $D=5$ dimensions with a generalized Einstein-Hilbert action 
\begin{equation}
S_{\text{5D}} = \frac{1}{16 \pi \tilde{G}}\int d^4x dy
\sqrt{-\tilde{g}_{AB}} \tilde{R}+ S_{\rm matter}\,,
\end{equation}
where $y$ indicates the compactified dimension, while  $\tilde{G}$, $\tilde{R}=\tilde{g}_{AB} \tilde{R}^{AB}$, and $ \tilde{g}_{AB}$ denote    the five-dimensional gravitational constant,  Ricci scalar, and metric tensor, respectively. Note also that $S_{\rm matter}$ gives the contribution of matter fields and includes a contribution coming from the complex scalar field $\tilde{\phi}$ with a self-interaction potential $V(\tilde{\phi})$. The latter can be written in  terms of the four-dimensional spacetime metric $g_{\mu \nu}$, the scalar field $\phi$ and the gauge field $A_{\mu}$ as
\begin{equation}
  \tilde{g}_{AB} =\left( \begin{matrix}
g_{\mu\nu} + \phi^2 A_\mu A_\nu & \phi^2 A_\mu \\[2mm]
\phi^2 A_\nu & \phi^2
\end{matrix}\right).
\end{equation}

It is known that after performing the dimensional reduction in the case $D=5$, we get a theory similar to a scalar-tensor model with an additional gauge field. By rescaling the field strength as $F_{\mu \nu}=\phi^{-3/2} \tilde{F}_{\mu\nu}$, and renaming $G=G_N/\phi$,  we are led to
\begin{align}
S_{\text{4D}} =& \int d^4x\,\sqrt{-g} \left( \frac{R}{16 \pi G}
- \frac{1}{4} \tilde{F}^{\mu \nu }\tilde{F}_{\mu
\nu }  - \frac{1}{2}\partial_\mu \tilde{\phi}\partial^\mu \tilde{\phi}-V(\tilde{\phi})\right),
\label{4D-action}
\end{align}
where Newton's constant has been expressed as $G=G_N (1+\alpha)$,  $\alpha$  being a parameter. 

One crucial point of KK gravity  is the analogy with the process of superconductivity, where photons become massive via the spontaneous symmetry breaking mechanism \cite{Jusufi:2024utf}. 
This involves the complex scalar field $\tilde{\phi}$, which is minimally coupled to the gauge field $A_\mu$ with gauge coupling constant $g$. Without going into details (see Ref. \cite{Jusufi:2024utf}), we rewrite the  scalar field  as $\tilde{\phi} = \tilde{\phi}_0e^{i\chi},$ where $\tilde\phi_0$ is the non-vanishing vacuum expectation value, and then 
upon employing the Anderson-Higgs mechanism, one obtains that the gauge invariant Lagrangian in KK model can be written as 
\begin{equation}
\mathcal{L} =\frac{1}{16 \pi G} R-\dfrac{1}{4} \tilde{F}^{\mu \nu }\tilde{F}_{\mu
\nu }  - \frac{1}{2}\mu^2 \tilde{A}^{\mu }\tilde{A}_{\mu }-V(\tilde{\phi}),
\label{KK-Lagr}
\end{equation}%
where the gauge boson mass reads as $\mu^2 \equiv g^2|\tilde\phi_0|^2$,  $V(\tilde{\phi})$ is the self-interaction potential of the scalar field $\tilde{\phi} $, and $\tilde{A}_\mu$ is connected to $A_\mu$ via the gauge transformation $\tilde{A}_\mu = A_\mu - (\partial_\mu\chi) /g$. 

The degrees of freedom can now be counted as follows. A massless graviton in $D=5$ dimensions has five degrees of freedom, given by the formula  $D(D-3)/2|_{D=5}=5$. Then, in $D=4$ dimensions we end up with two degrees of freedom for the massless spin-2 graviton, two for the massless spin-1 graviton, and one for the scalar graviton. 

The $U(1)$ symmetry breaking mechanism in superconductors is responsible for the occurrence of the mass term $\mu^2 \tilde{A}_\mu\tilde{A}^\mu/2$. This arises from the 
the degree of freedom of the complex scalar field $\tilde{\phi}$, which is  absorbed by the gauge boson. As a result, we have two degrees of freedom for the massless spin-2 graviton and three for the massive spin-1 graviton.

The variations of the gravitational Lagrangian with respect to the metric yields the Einstein field equations \cite{Jusufi:2024utf}
\begin{equation}
G_{\mu \nu}+\Lambda g_{\mu \nu}= 8 \pi G \left( T_{\mu \nu}^{\rm V}+T_{\mu \nu}^{\rm M}\right),
\end{equation}
with $T_{\mu \nu}^{\rm M}$ being the matter part of the stress-energy tensor, and $T_{\mu \nu}^{\rm V} $ the energy-momentum tensor for massive spin-1 graviton. It should be noted that, in the strong-gravity regime, we also expect a contribution from the coupling between the vector field and the spacetime background geometry, resulting in a correction term in the energy-momentum tensor, $T^{\rm correc.}_{\mu \nu}(g_{\mu \nu}, A_{\mu})$. However,  since in this work we focus on the weak-gravity regime, we will neglect such contributions. 

From the Lagrangian \eqref{KK-Lagr}, one finds that the motion of the vector field is governed by the  Proca equation 
\begin{equation}
\nabla_\mu \tilde{F}^{\mu\nu}-\mu^2 \tilde{A}^\mu = 0,
\end{equation}%
which by choosing the Lorenz gauge, $ \nabla_{\mu}\tilde{A}^{\mu}=0$, and after some calculations, leads to a wave equation for $\tilde{A}^{\mu}$: 
\begin{eqnarray}\label{waveequation}
\left(\Box-\mu^2 \right) \tilde{A}^{\mu}-g^{\mu\alpha}R_{\alpha\nu}\tilde{A}^{\nu}=0.
\end{eqnarray}
where the term with the Ricci tensor is negligible in the weak-field limit. This particle is usually referred to as the massive spin-1 dark graviton. Its role in the gravitational AB effect will be explored in the next sections. 

The massive spin-1 dark graviton  can couple to baryonic matter with a dimensionless coupling parameter $\alpha_B$. The resulting gravitational force between the baryonic matter fields stemming from the gauge boson exchange is described in terms of a Yukawa gravitational potential, which in the static and spherically symmetric case reads as \cite{Jusufi:2024utf}
\begin{equation}
\Phi_{\text{YU}}\left( r\right) = \alpha_B G_N M \frac{e^{- r/\lambda}}{r},
\label{Yukawa-potential-1}
\end{equation}
where  the length scale $\lambda$ (which in galactic scales is expected  to be of kpc order \cite{Jusufi:2024utf}) is related to the mass of the vector boson via the relation  $\mu=1/\lambda$.  Since  $\alpha_B$ should be positive, Eq. \eqref{Yukawa-potential-1} refers to a repulsive  gravitational interaction.  

The Yukawa potential \eqref{Yukawa-potential-1} modifies the usual Newtonian law of gravity and gives rise to a model that can explain the emergence of the dark sector,  both in galaxies and at cosmological scales,  as an apparent effect given by $ \Omega^{(1)}_{\rm \rm DM,0}=\hat{\alpha} \Omega_{\rm B,0}$,  followed by an additional contribution term $\Omega_{\rm DM,0}^{(2)}= \sqrt{2\,\alpha_B\,\Omega_{\rm B,0}  \Omega_{\Lambda,0}}$, where $\hat{\alpha} :=\alpha-\alpha_B$, while $\Omega_{\Lambda,0}$, $\Omega_{\rm DM,0}$, and  $\Omega_{\rm B,0}$ are the density parameters for dark energy, dark matter, and baryonic matter, respectively (see  Ref. \cite{Jusufi:2024utf} for further details).

The contribution due to the standard gravitational interaction mediated by the  massless spin-2 graviton is described in terms of the potential 
\begin{eqnarray}
\Phi_{\rm B}(r)=-\frac{GM}{r}=-\frac{G_N (1+\alpha)M}{r},
\label{barionic-potential}
\end{eqnarray}
which, unlike the Yukawa potential  \eqref{Yukawa-potential-1}, yields  an attractive force. Therefore, the total potential $\Phi_{\rm tot}(r)$ becomes
\begin{align}
\Phi_{\rm tot}(r) =&\Phi_{\rm B}(r)+ \Phi_{\text{YU}}(r)=-\frac{G_N M }{r}\left(1+\alpha-\alpha_B e^{-\frac{r}{\lambda}}\right).
\label{tot-potential}   
\end{align}
In general, we expect the parameters $\alpha, \alpha_B$ and $\lambda$ to vary and depend on the properties of individual galaxies and scales.

\section{Gravitational Aharonov-Bohm phase in Kaluza-Klein modified gravity} \label{Sec:Gravitational-AB-phase}

The gravitational  AB effect arises from the influence of a time-dependent gravitational potential $\Phi_{g}$ on a quantum system, such as an atomic or nuclear system. In general, the quantum-mechanical wave function of the system can be transformed via a phase as $\psi'(\mathbf{x},t)=e^{-i \varphi} \psi(\mathbf{x},t)$, where $\varphi$ can be written in terms of the action as 
\begin{eqnarray}
    \varphi=\frac{1}{\hbar}S.
\end{eqnarray}
In terms of the metric, which in KK theory is just the four-dimensional GR metric $ g_{\mu \nu}$, the action is given by \cite{Stodolsky1979}
\begin{eqnarray}
    S=-\frac{1}{m}\int g_{\mu \nu} p^{\mu} p^{\nu} d\tau,
\end{eqnarray}
where  $m$ is the mass of the particle or quantum system under consideration (which, as we will see,  experiences the AB effect), $\tau$  the proper time, and $p^{\mu}=m dx^{\mu}/d\tau$  the four-momentum. Considering perturbations around the flat  Minkowski metric $\eta_{\mu \nu}$, we can write $g_{\mu \nu}=\eta_{\mu \nu}+h_{\mu \nu}$, and thus we obtain two terms for the phase 
\begin{eqnarray}
    \varphi=-\frac{1}{\hbar}\int \eta_{\mu \nu} m\, v^{\mu} dx^{\nu}-\frac{1}{2 \hbar}\int h_{\mu \nu} p^{\mu} dx^{\nu},
\end{eqnarray}
where $v^{\mu}=dx^{\mu}/d t$. Our particular interest is  the second term, namely, the gravitationally induced phase 
\begin{eqnarray}
    \varphi_g=-\frac{1}{2 \hbar}\int h_{\mu \nu} p^{\mu} dx^{\nu}.
\end{eqnarray}
As pointed out before, in KK gravity, the presence of an extra dimension introduces an additional spin-1 gravitational field $\tilde{A}_\mu$. This affects the motion of test particles, leading to a modified  geodesic equation, which, in the weak-gravity regime, can be written as
\begin{equation}\label{geodesic-modified}
m\left(\frac{d^2 x^\mu}{d\tau^2} + \Gamma^\mu_{\nu\lambda} \frac{dx^\nu}{d\tau} \frac{dx^\lambda}{d\tau}\right) = f^{\mu},
\end{equation}
where $f^{\mu}=q_g{\tilde{F}^\mu}_{\;\;\nu}\, dx^{\nu}/d\tau $ represents a force analogous to the Lorentz force, but caused by the additional vector-gravitational interaction in KK gravity theory that acts on a test particle with mass $m$. Here, $q_g$ plays the role of the gravitational charge due to the vector field interaction. Henceforth, we take advantage of the Equivalence Principle to set $m=q_g$ (this will be justified below). We can thus say that  due to the additional interaction with $\tilde{A}_\mu$, an extra term in the phase appears:
\begin{eqnarray}
    \varphi_g=-\frac{1}{2 \hbar}\int h_{\mu \nu} p^{\mu} dx^{\nu}- \frac{m}{\hbar} \int \tilde{A}_{\mu} dx^{\mu},
\end{eqnarray}
where $\tilde{A}_{\mu}$   reads as,  for the static and spherically symmetric case,
\begin{eqnarray}
\tilde{A}_{\mu}=\left(\Phi_{\rm YU}, 0, 0, 0\right),
\end{eqnarray}
with $\Phi_{\rm YU}$ being the Yukawa potential introduced in Eq. \eqref{Yukawa-potential-1}. Now, bear in mind  that the Newtonian  scalar potential sourced by baryonic matter can be defined as $\Phi_B=- h_{00}/2$, then by retaining only the first-order terms in the perturbation, we get 
\begin{eqnarray}
    \varphi_g=\frac{m}{\hbar}\int \Phi_B dt+\frac{m}{\hbar} \int \Phi_{\rm YU}  dt,
\end{eqnarray}
where    $\Phi_B $ can be read off from Eq. \eqref{barionic-potential}. Thus, we can conclude that the gravitational potential induces a phase given by
\begin{equation}
\varphi_{g}(t)=\frac{m}{\hbar} \int_{0}^t \Phi_{g}(t') dt'.
\label{phase-formula-1}
\end{equation}
This relation is formally equivalent to the one in Ref. 
\cite{Chiao:2023ezj}, with the key difference being that in this paper the gravitational potential $\Phi_{g}$ includes both Newtonian and Yukawa-like corrections, as given by Eq. \eqref{tot-potential}.  This means that, in our framework, the gravitational phase can be expressed as
\begin{equation}
\varphi_{g}(t)=-\frac{m}{\hbar}\int_{0}^t \frac{G_N M}{r(t')}\left(1+\alpha -\alpha_B e^{-\frac{r(t')}{\lambda} }\right) dt'.
\label{AB-phase-integral-2}
\end{equation}

Following the setup in Ref. \cite{Chiao:2023ezj}, we will explore the gravitational AB effect by considering a quantum system in free fall around a gravitating body, such as a satellite orbiting the Earth. In this context, the quantum system refers to atomic or nuclear systems within the satellite, such as atomic clocks or cold atom interferometers, which exhibit quantum coherence and phase sensitivity to gravitational potentials. These systems provide a precise way to probe the gravitational AB effect and investigate potential deviations from GR. 

In deriving the gravitational AB phase in Eq. \eqref{AB-phase-integral-2}, we assume that the Equivalence Principle extends to a good precision in KK gravity, meaning that the inertial and passive gravitational masses effectively remain identical even in the presence of the additional spin-1 graviton and scalar field. However, in general, there is no guarantee that the Equivalence Principle remains true in KK. The reason is the presence of a spin-1 graviton, which couples to matter and can cause particles to slightly deviate from geodesic motion. However, in the case of the solar system experiments, these effects are expected to be very small and can be neglected. Thus, we expect the Equivalence Principle to hold in our setup. Let us point out that there are other studies in KK gravity, which show that the presence of extra dimensions does not necessarily lead to a violation of the weak equivalence principle \cite{PONCE_DE_LEON_2009}. Specifically, it has been shown that modifications to mass ratios in KK gravity would significantly affect Kepler's third law, Lagrange points, and orbital polarization in planetary systems. Observations of the Sun, Moon, Earth, and Jupiter impose limits on such deviations that are three to six orders of magnitude stronger than previous constraints, indicating that extra dimensions play a negligible role in solar system dynamics \cite{PhysRevD.62.102001}. Therefore, within the regime of our analysis, assuming equivalence between inertial and gravitational mass remains a reasonable and well-supported approximation.

The  radius $r(t)$ of the orbit followed by the system can be written as 
\begin{equation}
r(t)=\frac{r_{p}+r_{a}}{2}+\frac{r_{p}-r_{a}}{2}\cos(\Omega t) \equiv A+B\cos(\Omega t),
\label{orbit-radius}
\end{equation}
where $r_{a}$ and $r_{p}$ denote the apocenter and pericenter radii, respectively. If we suppose $A\gg B$, then $r(t)$ can be approximated as
\begin{align}
\frac{1}{r(t)}=\frac{1}{A+B\cos(\Omega t)}\approx \frac{1}{A}  \left[1-\frac{B}{A}\cos(\Omega t)\right],   
\end{align}
and hence from Eq. \eqref{tot-potential} we find that the total potential takes the form
\begin{align}
\Phi_{\rm tot} (t) \equiv& \Phi_{g}(t) \approx -\frac{G_N M}{A}\biggl[\left(1-\frac{B}{A}\cos(\Omega t)\right)   \left(1+\alpha-\alpha_B \, e^{-\frac{A+B\cos(\Omega t)}{\lambda}}\right) \biggr].
\end{align}
Since the mass of the spin-1 graviton is expected to be very small, the length-scale parameter $\lambda=1/\mu$ can be considered large. Therefore, we can employ the approximate result $e^{-\frac{A+B\cos(\Omega t)}{\lambda}} \approx 1 - \frac{A+B\cos(\Omega t)}{\lambda}$, and  the evaluation of the integral \eqref{AB-phase-integral-2} thus yields 
\begin{align}\notag
\varphi_{g} (t) \simeq&  -\frac{G_{N}mM(1+\hat{\alpha})t}{\hbar A}-\frac{G_{N}mM \alpha_B}{\hbar \lambda }\left(1-\frac{B^2}{2 A^2}\right)t +  \frac{G_{N}mMB(1+\hat{\alpha})}{\hbar A^2 \Omega} 
\nonumber \\
& \times  \left(1+\frac{B \alpha_B
\cos(\Omega t)}{2(1+\hat{\alpha})\lambda}\right) \sin(\Omega t),
\label{var-phi-g-1}
\end{align}
which can be split into a linear contribution in  $t$ and a sinusoidal term $\varphi_{g}^{\prime}(t)$ as 
\begin{align}
  \varphi_{g} (t)   \simeq   -\frac{G_{N}mM(1+\hat{\alpha})t}{\hbar A}-\frac{G_{N}mM \alpha_B}{\hbar \lambda }\left(1-\frac{B^2}{2 A^2}\right)t + \varphi_{g}^{\prime}(t),
\label{var-phi-g-2}
\end{align}
with 
\begin{align}
\varphi_{g}^{\prime}(t):=& \frac{G_{N}mMB(1+\hat{\alpha})}{\hbar A^2 \Omega} \left(1+\frac{B \alpha_B
\cos(\Omega t)}{2(1+\hat{\alpha})\lambda}\right)  \sin(\Omega t). 
\label{sinusoidal-term}
\end{align}
If we let  $\lambda \to \infty$ and $\alpha =\alpha_B=0$, then  formula \eqref{var-phi-g-1} readily gives
\begin{equation}
\varphi_{g}^{\rm GR}(t)\simeq -\frac{G_{N}m M\,t}{\hbar A}+\frac{G_{N}m M B \sin(\Omega t)}{\hbar A^2 \Omega},
\label{GR-phase-shift-1}
\end{equation}
in perfect agreement with the GR case studied in Ref.  \cite{Chiao:2023ezj}. 

As we shall see below, the terms linear in $t$ occurring in Eq. \eqref{var-phi-g-2} will  yield an overall shift in the base energy of the atomic system, while the sinusoidal contribution \eqref{sinusoidal-term}  gives rise to the gravitational  AB phase. We can demonstrate this result by solving the Schr\"{o}dinger equation for a quantum system subject to the time-dependent gravitational potential $\Phi_{g}(t)$ \cite{Chiao:2023ezj}
\begin{equation}
i \hbar \frac{\partial \psi}{\partial t}=H \psi=\left(H_{0}+m \Phi_{g}(t)\right) \psi, 
\label{schrodinger}
\end{equation}
 $H_0$ being the time-independent part of the Hamiltonian. Equation \eqref{schrodinger} can be solved via the separation-of-variables \emph{ansatz} 
\begin{align}
\psi(\mathbf{x}, t)=X(\mathbf{x}) T(t),   
\label{ansatz}
\end{align}
thereby yielding 
\begin{align}
i \hbar \frac{\partial \psi}{\partial t}  =  i \hbar X \frac{d T}{d t}=\left(H_{0}+m \Phi_{g}\right) X T  = &T H_{0} X+X\left(m \Phi_{g}\right) T,
\end{align}
which can be further arranged to obtain 
\begin{equation}
-m \Phi_{g}+i \hbar \frac{1}{T} \frac{d T}{d t}=\frac{1}{X} H_{0} X.
\label{schrodinger-2}
\end{equation}
The left-hand side of Eq. \eqref{schrodinger-2} involves functions depending solely on $t$, while  the right-hand side depends only  on $\mathbf{x}$, i.e., the equation takes the general form $f(t)=g(\mathbf{x})$. This means that it necessarily implies that each function is equal to a constant, i.e.,  $f(t)=g(\mathbf{x})=E$. This allows us to write the two separated equations 
\begin{align}
&-m \Phi_{g}+i \hbar \frac{d \ln T}{d t}=E, 
\label{indschrodinger} \\
&\quad H_{0} X=E X. 
\label{indschrodinger-1}
\end{align}
Setting $X=\Psi_{i}(\mathbf{x})$ and $E=E_{i}$, Eq. \eqref{indschrodinger-1} amounts to the eigenvalue problem of the unperturbed time-independent Hamiltonian $H_{0}$, i.e., $H_{0} \Psi_{i}(\mathbf{x})=E_{i} \Psi_{i}(\mathbf{x})$. On the other hand, integrating Eq.  \eqref{indschrodinger} over $t$ yields the following solution for $T(t)$:
\begin{align}
T(t) &=e^{-i \zeta t/\hbar}\,e^{-i \varphi_{g}^{\prime}(t)},
&
&
\label{T(t)}
\end{align}
where 
\begin{align}
\zeta:=&E_{i}+\frac{G _{N}mM(1+\hat{\alpha})}{A} + \frac{G_{N}mM \alpha_B}{ \lambda }\left(1-\frac{B^2}{2 A^2}\right).
\label{zeta-function}
\end{align}
Therefore, bearing in mind the \emph{ansatz} \eqref{ansatz} jointly with the relation $X(\mathbf{x})=$ $\Psi_{i}(\mathbf{x})$,  we find that the wave function for the Hamiltonian $H=H_{0}+m \Phi_{g}(t)$ reads as 
\begin{equation}
\psi_{i}(\mathbf{x}, t)=\Psi_{i}(\mathbf{x}) e^{-i \zeta t/\hbar}\,e^{-i \varphi_{g}^{\prime}(t)}.
\label{wave-function-2}
\end{equation}
Here, we note that the presence of the AB phase factor $\exp \left(-i \varphi_{g}^{\prime}(t)\right)$ does not alter  the probability density $\rho=\psi_{i}(\mathbf{x}, t)  \psi^{\star}_{i}(\mathbf{x}, t)$, which is thus not affected by the modified potential.

It readily follows from  Eq. \eqref{sinusoidal-term}, that the exponential term $\exp \left(-i \varphi_{g}^{\prime}(t)\right)$  occurring in the wave function \eqref{wave-function-2} takes the form 
\begin{align}
&e^{-i \varphi_{g}^{\prime}(t)} =\exp{\left(-i \frac{G_{N}mM (1+\hat{\alpha}) B}{\hbar A^2 \Omega} \sin (\Omega t)\right)}\exp{\left(-i \frac{G _{N}mMB^2
\alpha_B
}{4 \hbar A^2 \Omega \lambda } \sin (\Xi t) \right)}, 
\end{align}
where we have employed the trigonometric identity $2 \sin (\Omega t) \cos(\Omega t)=\sin(2 \Omega t)$ along with the definition $\Xi:=2\Omega$. In this way, invoking the  Jacobi-Anger expansion \cite{Cuyt2008}
\begin{align}
 e^{iz \sin \theta} &= \sum_{n=-\infty}^{+\infty}  J_n(z) e^{in\theta}, 
 \end{align}
 where $J_n(z)$ indicates the $n$-th Bessel function of the first kind satisfying $J_{-n}(z)= (-1)^n J_{n}(z)$, 
 we can write 
\begin{align}
&e^{-i \varphi_{g}^{\prime}(t)} =\sum_{n=-\infty}^{\infty}(-1)^{n} J_{n}\left(\frac{G_{N}mM(1+\hat{\alpha})B}{\hbar A^2 \Omega}\right) e^{i n \Omega t}  \sum_{k=-\infty}^{+\infty}(-1)^{k} J_{k}\left(\frac{G_{N}mM B^2 \textcolor{black}{\alpha_B}
}{4\lambda\hbar A^2 \Omega}\right) e^{i k \Xi t}.
\label{weighting}
\end{align}
Therefore, a comparison with Eqs. \eqref{zeta-function} and \eqref{wave-function-2}  shows that each energy level $E_{i}$ can be split into a multiplet $E_{i}^{(n)}$,  with
\begin{align}
E^{(n)}_{i}=& E_{i}+\frac{G_{N} Mm}{A}\left[1+\hat{\alpha}+\frac{\alpha_B A}{\lambda}-\frac{\alpha_B B^2}{2A\lambda}\right] 
\pm  (n+2k)\hbar \Omega.   \label{energy-split-formula}
\end{align}

We thus find from Eq. \eqref{energy-split-formula} that the energy levels are indeed corrected due to the modification of the law of gravity. Indeed, unlike in GR,  an additional energy splitting is induced by the parameters $\hat{\alpha}$ and $\alpha_B$. Furthermore, we identify another correction term proportional to $\hbar \Omega$. This splitting term is associated with new quantum numbers stemming from the interaction between baryonic matter and the spin-1 graviton, which introduces extra angular momentum states. The presence of the term with $k$ in the energy shift expression  \eqref{energy-split-formula}  suggests that the interaction between a massive graviton and a quantum system induces spin-dependent effects, analogous to the Aharonov-Casher effect in electromagnetism \cite{PhysRevLett.53.319}. In the Aharonov-Casher phenomenon, a particle's spin interacts with an external field, leading to a phase shift without experiencing a classical force. Similarly, in the presence of a massive graviton, the spin of a quantum system  can lead to energy splitting, reflecting a spin-orbit-type interaction within the gravitational sector. This implies that a massive graviton not only modifies the structure of the gravitational potential,  but it also introduces new quantum effects. 

Finally, it should be noted that in the special case where $\hat{\alpha} \to 0$ and $\lambda \to \infty$, we recover from Eq. \eqref{energy-split-formula} the result consistent with GR
\begin{eqnarray} \label{GR-result}
E^{{\rm GR} \,(n)}_{i}= E_{i}+\frac{G_{N} M m}{A}\pm  n \hbar \Omega,
\end{eqnarray}
which agrees with the outcome of Ref. \cite{Chiao:2023ezj}. 

\section{Energy and phase shift due to Kaluza-Klein modified gravity}\label{Sec:Energy-phase-shift}

In this  section, we will numerically estimate and further elaborate  our findings. 

In our analysis, one particularly important quantity is the difference between the GR and Yukawa energy levels
\begin{equation}
\Delta  E^{(n)}_{i}:= E^{(n)}_{i}-E^{{\rm GR} \,(n)}_{i},
\end{equation}
which using Eqs. \eqref{energy-split-formula} and \eqref{GR-result} yields the following result:
\begin{align}
\Delta  E^{(n)}_{i}=&\frac{G_{N} Mm }{A}\left[\hat{\alpha}+\frac{\alpha_B A}{\lambda}- \frac{\alpha_B B^2}{2A\lambda}\right] 
 \pm (2k)\hbar \Omega.
\end{align}
Let us now specialize to almost circular, low-Earth orbit, for which $A = 6.805 \times 10^{6}$ m, $B = -5 \times 10^{3}$ m (cf. Eq. \eqref{orbit-radius}). In addition, the value of $m$ depends on the quantum system being used for the measurements, and in the present paper we use electron mass, i.e.,  $m=9.1 \times 10^{-31} $ kg. Then, if we set $k=0$,  we get a shift in the energy expressed in eV units 
\begin{equation}
\Delta  E^{(n)}_{i}=(0.332\times 10^{-3} \hat{\alpha} +6.437 \times 10^{45}\hat{m}_g )\,\text{eV}, \label{energyshift}
\end{equation}
with $\hat{m}_g=\alpha_B m_g$, measured in kg units.  Another choice would involve a nuclear system  with mass $m=1.67 \times 10^{-27}$ kg (i.e., a neutron), as done in Ref. \cite{Chiao:2023ezj}. In that case we get
\begin{equation}
\Delta  E^{(n)}_{i}=(0.609 \hat{\alpha} +1.181 \times 10^{49}\hat{m}_g )\,\text{eV}. \label{energyshift-nuclear}
\end{equation}

\begin{figure}[]
\centering\includegraphics[scale=0.65]{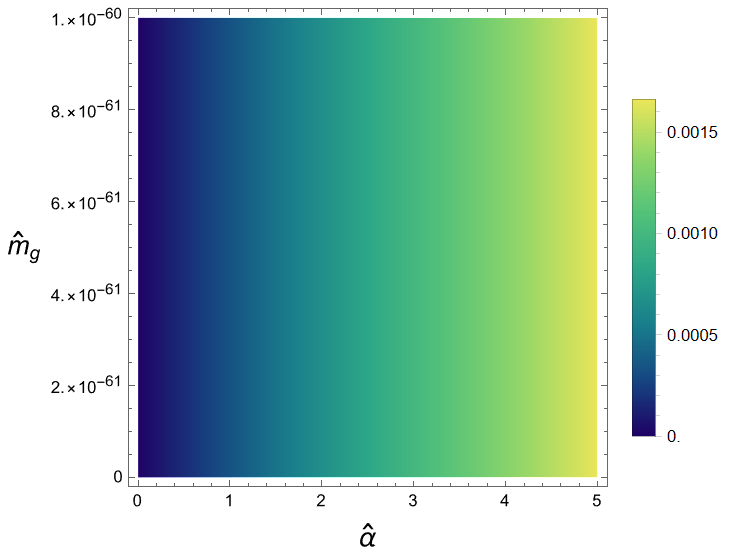}
\caption{Density plot for the energy shift for an atomic system measured in eV units (see Eq. \eqref{energyshift}) as a function of $\hat{\alpha}$ (no units) and $\hat{m}_g$, measured in kg units.}
\label{Fig-1}
\end{figure}

In what follows we are going to explain more about the expected range of the graviton mass. In Tables \ref{table-1} and \ref{table-2},  we present the numerical values for the energy shift using Eq. \eqref{energyshift}. We can see that with the decrease of the graviton mass, $\Delta  E^{(n)}_{i}$ attains smaller values as $\hat{\alpha}$ is fixed. On the other hand, if the graviton mass is fixed, then the increase of $\hat{\alpha}$ implies that $\Delta  E^{(n)}_{i}$ increases. We can deduce this fact also from  Fig. \ref{Fig-1}, where we give the density plot for the energy shift relation \eqref{energyshift} in terms of $\hat{\alpha}$ and $\hat{m}_g$. The typical graviton mass in galaxy scales is expected to be  $\hat{m}_g \sim 10^{-62}$ kg\footnote{This is consistent with the value of the graviton mass found from galactic rotating curves using a  total potential (Yukawa   plus the Newtonian contribution) similar to the one considered in Eq. \eqref{tot-potential}; see Ref.  \cite{J3} for further details.} and the energy shift for electron is of the order $\Delta  E^{(n)}_{i} \sim 0.16$ meV for specific value $\hat{\alpha}=0.5$. It is remarkable that our model is able to reproduce these expected results.  In addition, for the nuclear system, the effect is slightly stronger, namely from Eq. \eqref{energyshift-nuclear} we get $\Delta  E^{(n)}_{i} \sim 0.34$ eV using the same value $\hat{\alpha}=0.5$. This should not come as a surprise, as a quantum system with a larger  mass gives a stronger effect for the energy difference. However, we note that the a larger mass leads to smaller values of the maximum weighting for the Bessel functions  $J_n$ (cf.  Eq. \eqref{weighting}); we refer the reader to Ref. \cite{Chiao:2023ezj} for further details.
\begin{table}[h!]
    \centering
    \begin{tabular}{|c|c|}
        \hline
        \( \Delta  E^{(n)}_{i}[\rm eV]\) & $\hat{m}_g\ [\rm kg] $ \\ \hline
        6.437574524   &  $10^{-45}$   \\ \hline
        0.000230551     &  $10^{-50}$   \\ \hline
        0.000166178   & $10^{-55}$  \\ \hline
        0.000166177    & $10^{-60}$    \\ \hline
        0.000166177    & $10^{-65}$    \\ \hline
        0.000166177    & $10^{-68}$ \\ \hline
    \end{tabular}
    \caption{Numerical results for the energy shift given by Eq. \eqref{energyshift}, with a fixed parameter $\hat{\alpha}=0.5$ and  different values of  the graviton mass $\hat{m}_g$.}
    \label{table-1}
\end{table}
\begin{table}[h!]
    \centering
    \begin{tabular}{|c|c|}
        \hline
        \( \Delta  E^{(n)}_{i}[\rm eV]\) & $\hat{\alpha}\  $ \\ \hline
        $0.000033235514$   &  $0.1$   \\ \hline
        $0.000166177571$    &  $0.5$   \\ \hline
        $0.000332355142 $   & $1.0$  \\ \hline
        $0.000664710284 $  & $2.0$    \\ \hline
       $ 0.000997065426 $    & $3.0$    \\ \hline
        $0.001329420569 $   & $4.0$ \\ \hline
    \end{tabular}
    \caption{Numerical results for the energy shift given by Eq. \eqref{energyshift}, with a fixed mass $\hat{m}_g=10^{-62}$ kg and  different values of  the parameter $\hat{\alpha}$.}
    \label{table-2}
\end{table}

A possible experimental setup  for measuring the gravitationally induced energy differences \eqref{energyshift} and \eqref{energyshift-nuclear} might involve  the use of atomic clocks, as pointed out in Ref. \cite{Chiao:2023ezj}. Moreover, the predicted energy level splitting due to the gravitational AB effect in the presence of a Yukawa-like potential suggests an avenue for experimental detection via high-precision atomic interferometry and clock experiments. One particularly promising approach involves the use of the Space-Time Explorer and Quantum Equivalence Principle Space Test (STE-QUEST) mission \cite{Ahlers:2022stequest}, which is designed to test gravitational time dilation and equivalence principle violations using both atom interferometry and ultra-precise atomic clocks. In this setup, atomic wave packets experience different gravitational potentials along separate trajectories, accumulating a phase difference that is imprinted on the interference fringes. The modified gravitational potential in our model introduces an additional time-dependent phase shift, which could introduce an oscillatory modulation in the observed interference pattern, potentially serving as a signature of the gravitational AB effect.

Furthermore, since atomic clocks operate by measuring the frequency of well-defined atomic transitions, the predicted energy level splitting of  the order of $ 0.1 \, \text{meV}$ due to the spin-1 graviton would lead to a periodic modulation in atomic transition frequencies, potentially detectable through high-precision clock comparisons. If observed, such an effect would provide a direct test of modifications to gravity at microscopic scales, complementing astrophysical constraints from gravitational-wave detections \cite{PhysRevLett.121.251103}. This highlights the potential of STE-QUEST and similar future missions to probe quantum gravitational effects beyond GR.\\

We now turn our attention to the phase shift. Bearing in mind the expressions for the phase shift in the Yukawa-modified theory  and in GR given in Eqs. \eqref{var-phi-g-1} and \eqref{GR-phase-shift-1}, respectively, we can define the quantity 
\begin{equation}
\Delta \varphi_{g}(t) := \varphi_{g}(t) - \varphi_{g}^{\rm GR}(t).
\end{equation}
Then, we express the length scale $\lambda$ occurring in the Yukawa potential \eqref{Yukawa-potential-1} in terms of  the Compton wavelength, which can be related to the graviton mass $m_g$ through the formula $\lambda \approx \frac{\hbar }{m_g c}$. In this way, $\Delta \varphi_{g}(t)$ becomes  
\begin{align}
\Delta \varphi_{g} (t) \simeq &-\frac{G_N mM\hat{\alpha}\,t}{\hbar A} - \frac{G_N mM \textcolor{black}{\hat{m}_g} c\, t}{\hbar^2  }
 \left(1-\frac{B^2}{2 A^2}\right) +
\frac{G_N mM\hat{\alpha} B}{\hbar A^2 \Omega} 
\nonumber \\
&\times \left(1+\frac{B \textcolor{black}{\hat{m}_g} c \cos(\Omega t)}{2\hat{\alpha}\hbar }\right) \sin(\Omega t).
\label{delta-phi-g-1}
\end{align}
As $\hbar \to 0$, $\Delta \varphi_g$ blows up, indicating a failure of the classical intuition. 
Normally, quantum mechanics reduces to classical mechanics in the regime $\hbar \to 0$. However, in effects that are purely quantum, like AB interference, the limit $\hbar \to 0$ does not smoothly transition to a classical counterpart, but rather leads to divergent quantities. Since the AB effect relies on quantum interference, which disappears in the classical limit, the phase shift becomes ill-defined when  $\hbar \to 0$.

\begin{table}[h!]
    \centering
    \begin{tabular}{|c|c|}
        \hline
        \(  |\Delta \varphi_{g}(t)|\) [rad] & $\hat{m}_g\ [\rm kg] $ \\ \hline
         $5.284287686\times 10^{19}$  &  $10^{-45}$   \\ \hline
           $1.892341461\times 10^{15}$ &  $10^{-50}$   \\ \hline
         $1.363931616 \times 10^{15}$ & $10^{-55}$  \\ \hline
          $1.363926332 \times 10^{15}$ & $10^{-60}$    \\ \hline
            $1.363926332 \times 10^{15}$ & $10^{-65}$    \\ \hline
           $1.363926332\times 10^{15}$ & $10^{-68}$ \\ \hline
    \end{tabular}
    \caption{Numerical results for the phase shift given by Eq. \eqref{phaseshift}, with a fixed parameter $\hat{\alpha}=0.5$ and different values of the graviton mass.  }
    \label{table3}
\end{table}
\begin{table}[h!]
    \centering
     \begin{tabular}{|c|c|}
        \hline
        \(  |\Delta \varphi_{g}(t)|\) [rad] & $\hat{\alpha} $ \\ \hline
         $2.727852664\times 10^{14}$  &  $0.1$   \\ \hline
           $1.363926332\times 10^{15}$ &  $0.5$   \\ \hline
         $2.727852664 \times 10^{15}$ & $1.0$  \\ \hline
          $5.455705328 \times 10^{15}$ & $2.0$    \\ \hline
            $8.183557992 \times 10^{15}$ & $3.0$    \\ \hline
           $1.091141066\times 10^{16}$ & $4.0$ \\ \hline
    \end{tabular}
    \caption{Numerical results for the phase shift given by Eq. \eqref{phaseshift}, with a fixed  graviton mass $\hat{m}_g=10^{-62}$ kg and different values for parameter $\hat{\alpha}$.}
    \label{table4}
\end{table}

In almost circular, low-Earth orbit setup, the orbital period $t \approx 5400 \, \text{s}$, which corresponds to the approximate orbital period of satellites like the International Space Station. This gives an angular frequency $\Omega  = 2 \pi /t \approx 10^{-3} \, \text{rad/s} $. Using these numbers, from Eq. \eqref{delta-phi-g-1} we obtain for the atomic system
\begin{equation}\label{phaseshift}
   |\Delta \varphi_{g} (t)| \approx 2.727852 \times 10^{15} \hat{\alpha}+5.284151 \times 10^{64} \hat{m}_g.
\end{equation}

\begin{figure*}[]
\centering\includegraphics[scale=0.65]{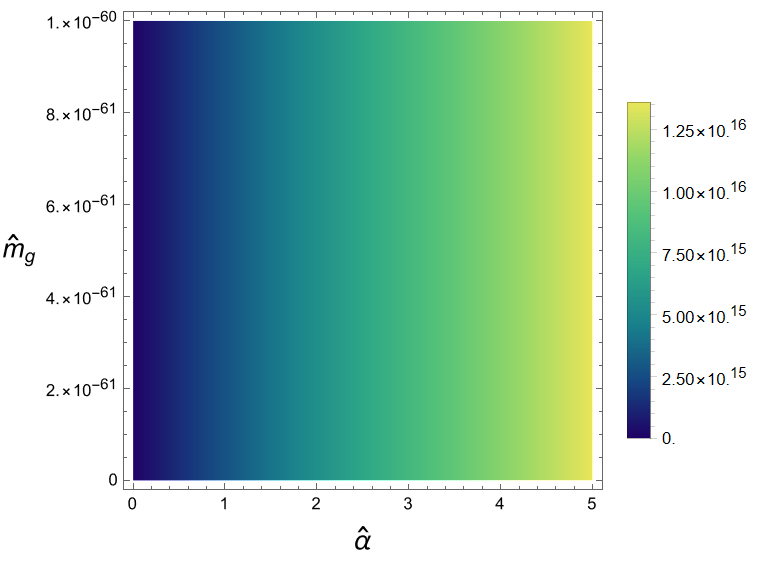}
\caption{Density plot for phase shift measured in rad units (see Eq. \eqref{phaseshift}) as a function of $\hat{\alpha}$ (no units) and $\hat{m}_g$, measured in kg units.}\label{fig2}
\end{figure*}

The values for the phase shift $|\Delta \varphi_{g} (t)|$ are displayed in the Tables \ref{table3} and \ref{table4}, while its density plot is showed in Fig. \ref{fig2}. We can see that $|\Delta \varphi_{g} (t)|$ decreases as the   graviton mass decreases while keeping $\hat{\alpha}$  fixed, and it increases when $\hat{\alpha}$ increases and $\hat{m}_g$ remains fixed.  Again, for galaxy scales we expect the graviton mass to be $\hat{m}_g \sim 10^{-62}$ kg, which corresponds to a  phase shift $|\Delta \varphi_{g} (t)| \sim 10^{15}$ rad.  
We note that the significant phase shift observed in our system is primarily attributable to two key factors: the large orbital radius and the extended time period. Together, these factors result in a greater accumulation of phase change.

\section{Aharonov-Bohm effect from other modified gravity theories}\label{Sec:AB-effect-other-theories}

Yukawa-type modifications to the gravitational potential emerge in various modified theories of gravity, including bimetric massive gravity \cite{Aoki:2016zgp}, Horndeski scalar-tensor theory \cite{Amendola:2019laa}, extended or modified $f(R)$ gravity \cite{Capozziello:2011et}, and models equivalent to GR  with graviton mass induced by dark energy \cite{J4, J5}. By examining the weak-field limit of extended gravity theories, one can show the emergence of the gravitational Yukawa-like potential \cite{Aoki:2016zgp,Amendola:2019laa,Capozziello:2011et,J4, J5} 
\begin{equation}
\Phi(r) = -\frac{G M}{r} \left(1+\alpha \,   e^{-\frac{r}{\lambda}}\right),\label{yukawamod}
\end{equation}
where, like before, $\lambda$ represents the range of interaction due to the massive graviton. We point out  that there is a sign difference in the Yukawa correction term compared to Yukawa potential in the KK model, given by Eq. \eqref{Yukawa-potential-1}. This is explained by the fact that in KK theory, the gravity force mediated by  the spin-1 graviton is  repulsive, while the force stemming from the massive scalar graviton in Eq. \eqref{yukawamod} is attractive. This additional force  can provide an explanation for dark matter. However, as already noted in Ref. \cite{Jusufi:2024utf}, KK theory exhibits a remarkable feature:  in the galactic center, the repulsive force from the spin-1 graviton is suppressed by an additional attractive contribution arising from corrections to Newton's constant due to the scalar field, resulting in an almost-Newtonian attractive force. In contrast,  in the galaxy's outer regions, the repulsive force fades, allowing the attractive force to dominate. This effect has been attributed to dark matter, making the model similar to scalar-vector-tensor gravity theory \cite{Moffat:2005si}.

Another important aspect of the Yukawa potential \eqref{yukawamod} concerns the spin nature of the massive graviton, which depends on the specific theory under consideration. For example, it is known that $f(R)$ gravity, under certain conditions, can be reformulated as an equivalent scalar-tensor gravity theory \cite{Capozziello:2011et}. In such a case, an additional degree of freedom naturally emerges in the form of a scalar particle that couples to matter gravitationally. Consequently, the theory contains a massless graviton, as in GR, along with a massive scalar particle, sometimes dubbed the scalaron or massive scalar graviton.  In other theories, such as  bimetric gravity \cite{Aoki:2016zgp} or more recent models where the graviton mass is induced by dark energy  \cite{J4, J5}, an additional degree of freedom arises from a massive spin-2 graviton. Despite these differences, the mathematical formulation of the potential remains similar in all these models.

Starting from Eq. \eqref{yukawamod} and performing  similar calculations as before, we obtain that the energy levels $\tilde{E}_{i}$ are  split into  multiplets $\tilde{E}_{i}^{(n)}$  as follows 
\begin{align}
  \tilde{E}^{(n)}_{i}=& \tilde{E}_{i}+\frac{G Mm}{A}\left[1+\alpha-\frac{\alpha A}{\lambda}+ \frac{\alpha B^2}{2A \lambda}\right] 
\pm  (n-2k)\hbar \Omega.
\end{align}
Therefore,  the difference between the Yukawa and GR energy levels reads as
\begin{align}
\Delta  \tilde{E}^{(n)}_{i}:=& \tilde{E}^{(n)}_{i}-E^{{\rm GR} \,(n)}_{i}
=\frac{G Mm}{A}\left[\alpha -\frac{\alpha A}{\lambda}+ \frac{\alpha B^2}{2A\lambda}\right] \mp  (2k)\hbar \Omega,
\end{align}
which upon setting $k=0$ and substituting the various  values   for the physical constants gives, for the atomic system, an energy shift in eV units 
\begin{equation}
\Delta  \tilde{E}^{(n)}_{i}=(0.332\times 10^{-3} \alpha-6.437 \times 10^{45}\hat{m}_g )\,\text{eV},
\label{energyshift-mod}
\end{equation}
while for the nuclear system we get 
\begin{equation}
\Delta  \tilde{E}^{(n)}_{i}=(0.609 \hat{\alpha}-1.181 \times 10^{49}\hat{m}_g )\,\text{eV},
\label{energyshift-modnuclear}
\end{equation}
with $\hat{m}_g=\alpha m_g$. Basically, the calculations show a similar result for the leading-order terms with the replacement $\alpha \to \hat{\alpha}$ (cf. Eqs. \eqref{energyshift} and \eqref{energyshift-nuclear}). However,  compared to  the energy shift relations  obtained in KK gravity, an  important sign difference appears in the higher-order factors of Eqs. \eqref{energyshift-mod} and \eqref{energyshift-modnuclear}. This variation can be traced to the nature of particle spin, specifically the spin-1 graviton in our case. As previously noted, in KK gravity the spin-1 graviton induces a repulsive force. However, within the interior of a galaxy, or at its center, this force can be suppressed by the additional attractive force from the scalar field encoded in Newton's constant. In the outer regions of the galaxy, the repulsive force diminishes due to its limited range, and the attractive force dominates. This effect is, in fact, attributed to dark matter.

Returning to the discussion of the energy shift, we note that higher-order corrections  are expected to be small, with only the first term playing a crucial role.  Using the expected value for the graviton mass, $\hat{m}_g  \leq  10^{-60}$ kg, the effect remains consistent with that in KK theory. To illustrate this, we can use the constraint for the Yukawa potential derived from the rotating curves of our galaxy \cite{J3}, where it is found that $\alpha \sim 0.4$ and the graviton mass $m_g \sim 10^{-62}$ kg. Using these values,  we obtain for the atomic system  $\Delta  \tilde{E}^{(n)}_{i}=0.00013$ eV=0.13 meV (cf. Eq. \eqref{energyshift-mod}), while for nuclear system we get $\Delta  \tilde{E}^{(n)}_{i}=0.24$ eV (cf. Eq. \eqref{energyshift-modnuclear}).  This shows that the effect remains at the meV or eV scale, depending on the system, as predicted by KK theory. \\

\section{Discussion and conclusions}\label{Sec:Conclusions}

Traditional searches for dark matter typically rely on the direct or indirect detection of its particles, but   have, to date, led to negative results. In this paper, we have proposed an alternative approach to inferring dark matter evidence through modified gravity. Specifically, we have investigated the possibility of extracting dark matter signatures using the gravitational AB effect in the context of KK gravity. This model is particularly intriguing as it provides a consistent description of the dark sector without invoking fundamental dark matter particles. In this framework, the gravitational potential exhibits Yukawa-like corrections that  introduce modifications to the gravitational AB phase due to the presence of a scalar spin-0 graviton  and a massive spin-1 graviton,  alongside the usual massless spin-2 graviton. Moreover, the analysis of the quantum-mechanical wave function  reveals that that Yukawa contributions induce a splitting in the energy levels of a  system,    which is related to a new quantum number tied to the quantum nature of the gravitons. 

In our analysis, we have provided estimates for the  graviton mass in galaxy-scale phenomena. We have found that a typical value is  $\hat{m}_g \sim 10^{-62}$ kg, which  leads to an energy shift of the order of meV for an atomic system and eV for a nuclear system. It is thus natural to ask whether such a result could be experimentally verified. Although this open  question  lies beyond the scope of this paper,  a potential experimental setup to test or distinguish these energy differences, particularly for the gravitationally induced effect, could involve the use of high-precision atomic interferometry and atomic clocks.

It is reasonable to expect that if massive spin-1 gravitons exist, they should leave some imprints in the spectrum of gravitational waves. This conclusion ties in with the recent findings of the NANOGrav collaboration, which  have disclosed potential contributions of massive gravitons to gravitational waves in the nHz frequency range \cite{NANOGrav:2023gor}. Specifically, there exists a minimal frequency for the gravitational-wave signal due to the massive graviton, which can be computed using the relation  $f_{\rm min} \simeq \frac{m_g c^2}{2 \pi \hbar}$,  measured in   \text{Hz}. For the estimated graviton mass used in this  paper, $m_g \sim 10^{-62}$ kg, we get $f_{\rm min} \sim 10^{-12}$ Hz. Remarkably, this aligns with the predictions presented in Ref. \cite{J3}, as well as with those from Einstein-Cartan theory, as explored in Refs. \cite{Battista2022a,Battista2022b,Battista2023a,DeFalco2023b}. Moreover,  it is likely that, on galactic scales, the constraint for the graviton mass can be improved, say to $m_g \sim 10^{-59}$ kg. In this case, we would get a minimal frequency $f_{\rm min} \sim 10^{-9}$ Hz $\sim$ 1 nHz. However, it should be noted that constraining the graviton mass can be challenging for two reasons: the mass may depend on the environment (such as the type of galaxy), and there may be a quantum-mechanical limitation in measuring the mass due to the uncertainty in momentum and position.

\section*{Acknowledgements}
EB acknowledges the support of  Istituto Nazionale di Fisica Nucleare (INFN), {\it Iniziative Specifiche} MOONLIGHT2.

\bibliographystyle{JHEP}
\bibliography{references}{}

\end{document}